\documentclass[aps,pre,twocolumn,showpacs,superscriptaddress]{revtex4}
\usepackage{graphicx}
\usepackage{bm}
\bibstyle{apsrev.bib}

\begin{document}
\title{Inter-strand distance distribution of DNA near melting}
\author{M. Baiesi}
\affiliation{INFM-Dipartimento di Fisica, Universit\`a di Padova,
I-35131 Padova, Italy.}
\author{E. Carlon}
\affiliation{Theoretische Physik, Universit\"at des Saarlandes,
D-66041 Saarbr\"ucken, Germany}
\author{Y. Kafri}
\affiliation{Department of physics of complex systems, Weizmann
Institute of Science, Rehovot, Israel 76100.}
\author{D. Mukamel}
\affiliation{Department of physics of complex systems, Weizmann
Institute of Science, Rehovot, Israel 76100.}
\author{E. Orlandini}
\affiliation{INFM-Dipartimento di Fisica, Universit\`a di Padova,
I-35131 Padova, Italy.}
\author{A. L. Stella}
\affiliation{INFM-Dipartimento di Fisica, Universit\`a di Padova,
I-35131 Padova, Italy.} \affiliation{Sezione INFN, Universit\`a di
Padova, I-35131 Padova, Italy.}

\date{November 12, 2002}

\begin{abstract}
The distance distribution between complementary base pairs of the
two strands of a DNA molecule is studied near the melting
transition. Scaling arguments are presented for a generalized
Poland-Scheraga type model which includes self-avoiding
interactions. At the transition temperature and for a large
distance $r$ the distribution decays as $1/r^{\kappa}$ with
$\kappa=1+(c-2)/ \nu$. Here $\nu$ is the self-avoiding walk
correlation length exponent and $c$ is the exponent associated
with the entropy of an open loop in the chain. Results for the
distribution function just below the melting point are also
presented. Numerical simulations which fully take into account the
self-avoiding interactions are in good agreement with the scaling
approach.

\end{abstract}


\maketitle

\section{Introduction}
Melting or denaturation of DNA, whereby the two strands of the
molecule unbind upon heating, has been a subject of interest for
several decades. In experiments carried out since the sixties,
calorimetric and optical melting curves have yielded information
on the behavior of the order parameter (fraction of bounded
complementary base pairs) near the transition \cite{wart85}. This
parameter gives a global measure for the average degree of opening
of the molecule. With the recent advent of novel experimental
techniques which allow for single molecule manipulations, it has
become possible to obtain more detailed information on the
microscopic configurations of fluctuating DNA. For example, the
time scale of opening and closing of loops of denaturated segments
and some information about their steady-state distribution may be
obtained by fluorescence correlation spectroscopy techniques
\cite{bonn98}. Additional information is also gained by studies of
the response of the molecule to stretching, unzipping and
torsional forces \cite{DNA1,DNA2,DNA3,DNA4,DNA5}.

Theoretically the melting transition has been studied within two
main classes of models. The first, which we refer to as
Poland-Scheraga (PS) type models
\cite{zimm,pola66,pola66b,fish66}, considers the molecule as being
composed of an alternating sequence of double-stranded segments
and denaturated loops. Within the model, weights are assigned to
bound segments and unbound loops from which the nature of the
transition may be deduced. In a second approach which has been
employed to study the melting transition \cite{peyr89}, the DNA is
considered as a directed polymer (DP). Here the two strands are
described as directed random walks and they interact through a
short-range attractive potential. Using a transfer matrix method
the melting transition may be studied.

Within the directed polymer approach the distance distribution of
complementary base pairs is readily calculable. However, realistic
geometrical restrictions (such as self-avoiding interactions) are
not taken into account due to the over simplifying directed
polymer description. On the other hand, within the PS type models
geometrical restrictions may be accounted for more realistically.
It has recently been demonstrated \cite{kafr00,kafr02,kafr02B}
that a generalization of this model which includes the repulsive
self-avoiding interactions between the various segments of the DNA
chain may be analyzed. This is done using a scaling approach for
general polymer networks introduced by Duplantier
\cite{dup1,dup2}. The results for the nature of the transition and
for the loop-size distribution are in very good agreement with
recent numerical studies \cite{caus00,carl02,baie02}. In the PS
type models the order parameter and the loop size distribution
near the transition are readily calculable. However, as defined,
these models do not yield the inter-strand distance distribution.
It would be interesting to generalize the scaling picture of the
PS type models in order to study the inter-strand distance
distribution close to the melting point.

In this Paper we study the distance distribution between
complementary base pairs of the two strands within a PS approach.
We derive scaling results valid both at and below the melting
temperature and verify their validity by extensive numerical
simulations of a model on a lattice which fully embodies excluded
volume interactions.

The Paper is organized as follows: In Section II we derive a
scaling picture for the inter-strand distance probability
distribution both at and below the melting point. Scaling
relations linking the exponents of the loop size and distance
distributions are provided. The results of numerical studies of
the distribution functions confirming the scaling picture are
given in Section III. The main results are summarized in Section
IV.

\section{Scaling approach}

We start by briefly reviewing the main results of the PS approach.
Within the framework of these models one assigns length dependent
weights to both bound and unbound segments. A bound segment is
energetically favored over an unbound segment, while an unbound
segment (loop) is entropically favored over a bound one. A bound
segment of length $l$ is assigned a weight $w^l$, where
$w=\exp(-E_0/T)$, $E_0$ is the base pair binding energy, $T$ is
the temperature and the Boltzmann constant $k_B$ is set to $1$.
Here it is assumed that only complementary base pairs interact
with each other. The binding energy $E_0$ is taken to be the same
for all base pairs. An unbound segment (loop) of length $l$ is
assigned a weight
\begin{equation}
\Omega(l)=\frac{s^l}{l^c}, \label{eq:loop_entropy}
\end{equation}
where $s$ is a non-universal geometrical constant and $c$ is an
exponent which is determined by some universal properties of the
loop configurations. The nature of the melting transition depends
on the value of the exponent $c$ \cite{pola66b}. For $c \leq 1$
there is no transition, for $1<c \leq 2$ the transition is
continuous, while for $c>2$ the transition is first order.

Early works \cite{pola66b} have evaluated the exponent $c$ by
enumerating random walks which return to the origin yielding
$c=d/2$ in $d$ dimensions. The inclusion of excluded volume
interactions within a loop gives $c=d \nu$ \cite{fish66}, where
$\nu$ is the correlation length exponent of a self-avoiding random
walk. Here the self-avoiding interactions between a loop and the
rest of the chain are neglected. Both these estimates predict a
continuous transition ($c <2$) for any $d<4$. Numerical
simulations of chains of length of up to $3000$ where
self-avoiding interactions have been fully taken into account
suggested that in fact the transition in the infinitely long chain
limit is first order \cite{caus00}. Recently it has been suggested
\cite{kafr00,kafr02,kafr02B} that excluded volume interactions
between a loop and the rest of the chain may be taken
approximately into account using results for polymer networks of
arbitrary topology \cite{dup1,dup2}. It has been shown that for
loops much smaller than the chain length the entropy of the loop
has the same form as in Eq.~(\ref{eq:loop_entropy}), but with $c=d
\nu - 2 \sigma_3$. Here $\sigma_3$ is an exponent associated with
an order three vertex configuration defined and evaluated in
\cite{dup1,dup2}. In $d=3$ the exponent $c$ may be estimated to be
$c \simeq 2.11$. Since $c>2$ the transition is first order. Within
the PS type models the weight of a loop of size $l$, $P_{\rm
loop}(l)$ is given by
\begin{equation}
P_{\rm loop}(l)\sim \frac{e^{- l / \xi_l}}{l^c} \;, \label{eq:loopdist}
\end{equation}
where $\xi_l \sim |T-T_M|^{-1/(c-1)}$ for $1<c \leq 2$ and $\xi_l
\sim |T-T_M|^{-1}$ for $c>2$. Here $T_M$ is the melting
temperature. In a recent numerical study \cite{carl02} the loops
size distribution at the melting transition has been evaluated for
chains of length up to $200$ where self-avoidance is fully taken
into account. These simulations yield $c \approx 2.10(4)$ in good
agreement with the theoretical estimate.

We now use a scaling approach to study the complementary base-pair
distance distribution $P_{\rm dist}(r)$. The probability that,
within a loop of size $2l$, two complementary base-pairs are
separated by $\vec{r}$, scales as
\begin{equation}
P (\vec{r},l) = \frac{1}{l^{d \nu}} f \left( \frac{r}{l^\nu}
\right) \;, \label{eq:Prl}
\end{equation}
where $r=\vert \vec{r} \vert$ and $f$ is a scaling function. To
obtain $P_{\rm dist}(r)$ we integrate over the contribution of all
loops and over the angular degrees of freedom $d \omega$:
\begin{equation}
P_{\rm dist}(r) \sim \int_{0}^{\infty} dl \ P_{\rm loop}(l) \int d
\omega \ r^{d-1} \ l P(\vec{r},l) \;.\label{eq:disint}
\end{equation}
Note that the contribution of each loop is $l P (\vec r,l)$ since
each loop contains $l$ matching pairs and thus contributes $l$
times its average distance to the average of $P_{\rm dist}(r)$.
Inserting Eqs.~(\ref{eq:loopdist}) and (\ref{eq:Prl})
into Eq.~(\ref{eq:disint}),
one finds
\begin{equation}
P_{\rm dist}(r)  \sim \int_{0}^{\infty} dl \frac{e^{-l/\xi_l}}{l^{c}}
\frac{1}{l^{\nu-1}}\left(\frac{r}{l^\nu} \right)^{d-1} f \left(
\frac{r}{l^\nu} \right).
\label{eq:Pdist}
\end{equation}
At the transition one has $\xi_l^{-1} = 0 $, and the integral
scales with $r$ as
\begin{equation}
P_{\rm dist}(r,\xi_l^{-1}=0) \sim \frac{1}{r^{\kappa}} \;,
\end{equation}
where
\begin{equation}
\kappa=1+(c-2)/\nu\;. \label{eq:kappa}
\end{equation}
The estimated values for the exponents $c \simeq 2.11$ and
$\nu=0.588$ in $d=3$ yield $\kappa \simeq 1.19$.

Next we consider the distance distribution below the transition
where $\xi_l^{-1}>0$. Simple scaling analysis can not be carried
out and one has to take a specific form for the function $f$. A
general argument due to Fisher~\cite{FisherSA} for the end to end
distance of a self-avoiding walk yields the following form of
$f(x)$ for $x \gg 1$
\begin{equation}
f(x)\sim x^{\mu} \exp(-Dx^{\frac{1}{1-\nu}}) \;, \label{eq:dimu}
\end{equation}
where $\mu$ is a known exponent. This argument may be generalized
to consider the average distance between complementary pairs
within a loop, or a loop embedded in a chain, yielding the same
form but with a different exponent $\mu$ (to be discussed below).
Using this form the integral (\ref{eq:Pdist}) may be evaluated
using a saddle-point approximation. This gives
\begin{equation}
P_{\rm dist}(r)\sim \frac{\exp(- r/\xi_r)}{r^\eta} \;,
\label{eq:r_prob_lowT}
\end{equation}
for $r \gg \xi_r$, with
\begin{equation}
\eta=c-1/2-(1-\nu) (\mu +d) \;. \label{eq:eta}
\end{equation}
The characteristic distance $\xi_r$ is related to the length
$\xi_l$ by
\begin{equation}
\xi_r \propto \xi_l^\nu\;, \ \ \ \ \ {\rm for}\ \xi_l\to\infty
\label{eq:rel_xil_xir}
\end{equation}
so that $ \xi_r\propto |T-T_M|^{-\nu/(c-1)}$ for $1<c \leq 2$ and
$\xi_r \propto |T-T_M|^{-\nu}$ for $c>2$. In our case $c\simeq
2.11$ and therefore we expect $ \xi_r^{-1} \propto |T-T_M|^{\nu}$.

Within the approach introduced in \cite{kafr00} the exponent $\mu$
in the distribution function (\ref{eq:dimu}) should be evaluated
by considering the average inter-strand distance in a loop
embedded in a chain. Here for simplicity we adopt the approach of
Fisher \cite{FisherSA} and consider the exponent $\mu$ in the
inter-strand distance distribution within an isolated loop. Thus
the effect of self-avoiding interactions between the loop and the
rest of the chain on the exponent $\mu$ is not taken into account.
The calculation is rather lengthy and it is outlined in Appendix
A. The resulting exponent is found to be
\begin{equation}
\mu=\frac{1}{1-\nu} \left( 1/2 + 2d(\nu-1/2)-\gamma \right) \;,
\label{eq:mu}
\end{equation}
where $\gamma$ is the exponent associated with the number of
configurations of a random walk of length $L$ as given by $s^L
L^{\gamma-1}$. Thus, for a random, non self-avoiding loop where
$\gamma=1$ and $\nu=1/2$ one has $\mu=-1$ for any $d$. On the
other hand for a self-avoiding loop in $d=3$, where $\gamma
\approx 1.18$ one has $\mu=-0.37$. An estimate for $\eta$ may be
obtained by using the $c$ value of an isolated loop (namely $d
\nu$) in equation (\ref{eq:eta}) together with the above value of
$\mu$ to yield $\eta \approx 0.18$. It would be of interest to
derive an expression for $\mu$ in the case of a loop embedded in a
chain in order to fully take into account the effect of
self-avoiding interactions.

It is instructive to compare these results with the distance
distributions obtained within the DP approach. The exponent $c$
characterizes the number of directed walks which return to the
origin for the first time. This is known to be given
by~\cite{redner} $c=2-d/2$ for $d < 2$ and $c=d/2$ for $d>2$. In
$d=2$ there are logarithmic corrections so that the number of
configurations behaves as $s^l/(l \ln^2 l)$. Thus, one expects a
continuous melting transition for $d<4$ and a first order phase
transition for $d>4$. Clearly, the correlation length exponent
satisfies $\nu=1/2$. Using these results one obtains at
criticality
\begin{eqnarray}
\kappa_{DP}&=& d-3 \;\; {\mbox {\rm for}} \;\; d>2  \label{eq:kDP1}\\
&=& 1-d \;\; {\mbox {\rm for}} \;\; d<2 \;. \label{eq:kDP2}
\end{eqnarray}
Eqs.\ (\ref{eq:kDP1})-(\ref{eq:kDP2}) are
 in agreement with calculations using a transfer
matrix method for the DP model \cite{lipowsky}. Below criticality
our results predict that the distance distribution decays
exponentially with $r$ with $\xi_r \propto |T-T_M|^{-1/\vert 2-d
\vert}$ for $d<4$ and $\xi_r \propto |T-T_M|^{-1/2}$ for $d>4$.
Also, using $\mu=-1$ for the DP model one has $\eta=0$. These
results are again in agreement with known results for the DP model
\cite{lipowsky}.

\section{Numerical simulations}

In order to test the predictions of this scaling picture, we carried out
 extensive numerical simulations of the loop size and inter-strand
distance distributions of a model of fluctuating DNA \cite{caus00,carl02}.
The DNA strands are represented by two self-avoiding walks of
 length $N$ on a cubic lattice.
The numerical simulations are carried out by using the
pruned-enriched Rosenbluth (PERM) Monte Carlo method
\cite{gras97}, which has already been employed in recent studies
of DNA denaturation \cite{caus00,baie02}. This method generates
DNA chain configurations by a growth algorithm at fixed $T$. Each
configuration consists of two complementary sequences of $N$ unit
steps between nearest neighbor sites of the lattice, both
 starting from a common origin.
Self avoidance is achieved by forbidding overlapping of sites.
This constraint is relaxed only to introduce an interaction
between complementary sites (with the same index along the two
strands), which are allowed to overlap, with an energy gain
$E_0=-1$. In this way the total number of these contacts gives the
energy gain, $-E$, and the Boltzmann weight of a DNA configuration
is $\exp(E/T)$. In order to recover  the equilibrium distribution
in the simulation, one has to assign  a suitable weight for each
growth step of the chain \cite{gras97,rose55}. In the present
work we have modified the growth rules in order to achieve a
better performance at lower $T$, where ordinary PERM yields slower
convergence. In the usual PERM rules, long open segments which
have low weights at low $T$ are generated, and they are thus often
pruned. This makes it difficult to generate sufficiently long and
loop-rich chains by this procedure. In order to avoid this problem
we have introduced a small bias for the growing ends to recombine.
This bias is compensated by a suitable reweighting of the
generated chain, to yield to correct equilibrium distribution. The
results of this study, which are described below, corroborate the
scaling picture introduced above.

\begin{figure}
\includegraphics[angle=-90,width=8.4cm]{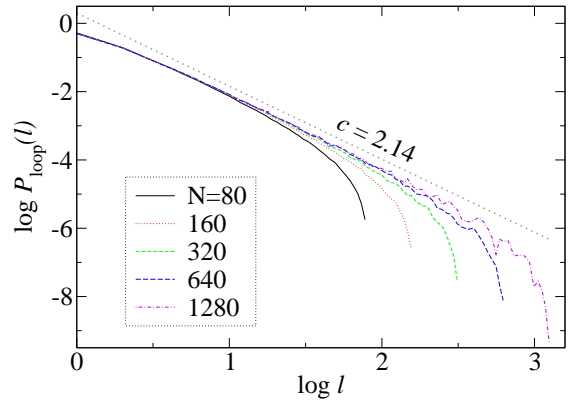}
\caption{Log-log plots of $P_{\rm loop}(l)$ at $T_M=0.7455$ for
several chain length $N$. One can identify a linear region (whose
range increases with $N$) with slope $-c = -2.14(4)$ (dotted
line). \label{fig:01}}
\end{figure}

We start by first considering the loop size distribution at the
melting temperature $T_M=0.7455$. This distribution has been
studied in the past for chains of length up to $N=320$ monomers
\cite{carl02,baie02}. In Figure \ref{fig:01} we present the
results for chains of length up to $N= 1280$. We find $c=2.14(4)$,
which is in good agreement with the analytical estimate
$c\simeq2.11$ and the previous numerical estimates obtained from
simulations of shorter chains.

\begin{figure}
\includegraphics[angle=-90,width=8.4cm]{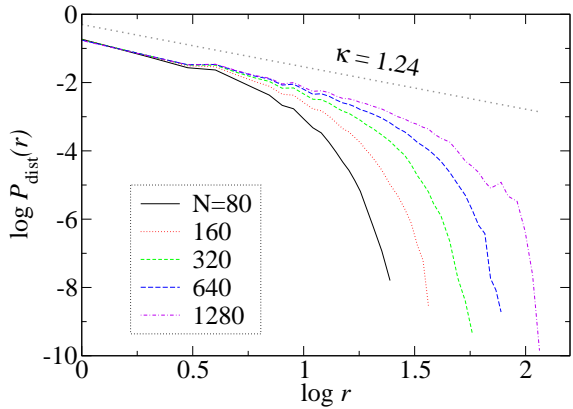}
\caption{Log-log plots of $P_{\rm dist}(r)$ at $T_M$ for several
chain length $N$. The slope $-\kappa = -1.24$, derived from
Eq.~(\ref{eq:kappa}) and plotted as a dotted line in the log-log
scale, is consistent with the trend developing in the
distributions of longer chains. \label{fig:02}}
\end{figure}

\begin{figure}[btp]
\includegraphics[angle=-90,width=8.4cm]{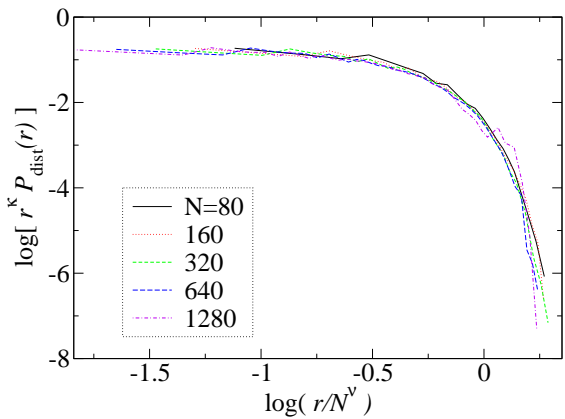}
\caption{Collapse plot of $P_{\rm dist}(r)$ according to a scaling
form $P_{\rm dist}(r,N) \simeq r^{-\kappa} g(r/N^\nu)$, where $\nu
\simeq 0.59$.
 \label{fig:03}}
\end{figure}

The complementary-pair distance distribution at the melting point
is plotted in Figure~\ref{fig:02} for systems of size up to
$N=1280$. We find that the decay exponent is given by
$\kappa=1.24(7)$ which is the expected value from the scaling
relation (\ref{eq:kappa}) given the measured value of $c$. A
direct estimate of $\kappa$ from the data is not easy, since the
power law behavior has a cut-off at values of $r$ which are much
smaller than those for the $l$ distribution. However, the
algebraic decay of $P_{\rm dist}(r)$ is confirmed by the good
collapse plot shown in Fig.~\ref{fig:03}.

\begin{figure}[tb]
\includegraphics[angle=-90,width=8.4cm]{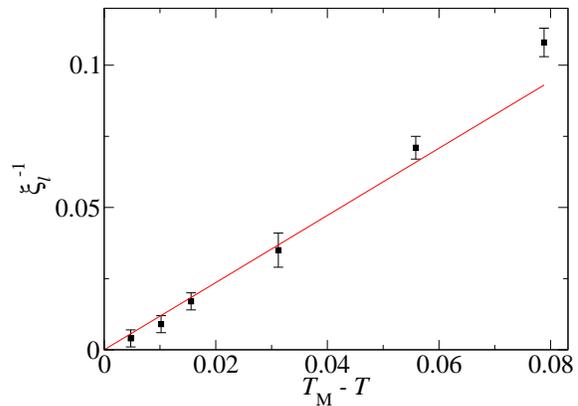}
\caption{The characteristic length $\xi_l^{-1}$ as a function of
$|T-T_M|$, for $T<T_M$. \label{fig:04}}
\end{figure}

We now consider the distribution functions below the melting
temperature and study the behavior of the length scales $\xi_l$
and $\xi_r$. Motivated by the asymptotic form (\ref{eq:loopdist})
for the loop size distribution we extract $\xi_l$ by fitting $\ln
P_{\rm loop}(l)$ to the form
\begin{equation}
y_0 -  l/\xi_l - c \ln l\,.
\label{eq:fit_xil}
\end{equation}
where $y_0$ is a constant. This fit is carried out for several
values of the temperature near $T_M$ using $c=2.14$. For each
temperature $T$, the values of $\xi_l$ is obtained at different chain
lengths $N$ and is then extrapolated to the limit $N \to \infty$.
The resulting temperature dependence of the extrapolated
$\xi_l$ is displayed in
Figure \ref{fig:04}. Indeed, the expected linear dependence of
$\xi_l^{-1}$ on the temperature difference $\vert T -T_M \vert$ is
observed near $T_M$.

\begin{figure}[tb]
\includegraphics[angle=-90,width=8.4cm]{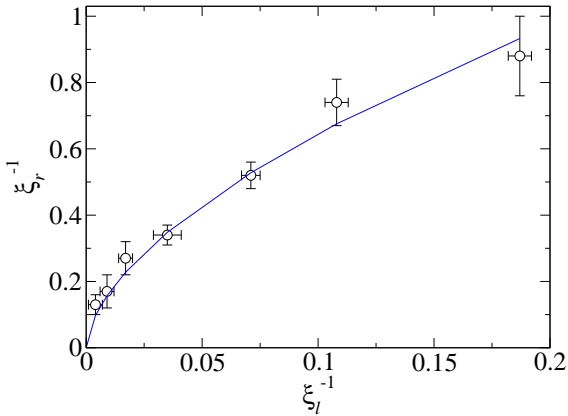}
\caption{Parameter $\xi_r^{-1}$ as a function of $\xi_l^{-1}$, for
$T<T_M$. Errors are indicated. For each $T$, we evaluate $\xi_r$
by a non-linear fit of $\ln P_{\rm dist}(r)$ of the form
(\ref{eq:fit_xir}). The solid line is a fit using the form
(\ref{eq:rel_xil_xir}). \label{fig:05}}
\end{figure}

In order to obtain $\xi_r$ we carried out a similar fit of  $\ln
P_{\rm dist}(r)$ to the form
\begin{equation}
y_1 -  r/\xi_r - \eta \ln r \,, \label{eq:fit_xir}
\end{equation}
where the constant $y_1$ and the parameter $\eta$ are left as free
parameters. Unfortunately, the numerical estimate of $\eta$ is
rather crude, yielding $0.5\lesssim \eta\lesssim 1.2$. In
Figure~\ref{fig:05} we present a plot of $\xi_r^{-1}$ as a
function of $\xi_l^{-1}$. This graph is consistent with the
expected form (\ref{eq:rel_xil_xir}).

Finally, we note that the scaling relations (\ref{eq:kappa}) and
(\ref{eq:eta}) are rather general and are not restricted to models
where self-avoiding interactions are taken into account. Recently,
Garel, Monthus and Orland (GMO) \cite{gare01} have introduced a
model for DNA denaturation where self-avoidance within each strand
is neglected while mutual avoidance is included. Each strand is a
simple random walk and thus $\nu=1/2$ for this model. Numerical
results obtained with the PERM method for this model in $d=3$
dimensions yield $c=2.55(5)$ and $\kappa = 2.1(1)$ \cite{baie02b}.
It is readily seen that these exponents satisfy the scaling
relation (\ref{eq:kappa}). In fact, for the GMO model one can also
develop a PS type of descrition \cite{baie02b} analogous to that
of Refs.\ \cite{kafr00,kafr02,kafr02B}, but based this time on a
block copolymer network picture \cite{baie02}. This description
gives  analytical $c$ estimates consistent with the numerical
results.

\section{Summary}
In this paper we studied the inter-strand distance distribution
for DNA at and near the melting point. A scaling analysis within
Poland-Scheraga type models where self-avoiding interactions are
taken into account is presented. A scaling relation is derived
(Eq. (7)) between the exponents $c$ and $\kappa$ which govern the
decay at melting of probability distributions of loop lengths and
of interstrand distances, respectively. Results of extensive
numerical simulations are found to be in agreement with the
scaling approach.

The DNA melting transition has been studied so far either with PS
type models or with the directed polymer approach. While in the
latter case $\kappa$ and $c$ are easily computable, in the PS
models the interstrand distance distribution, and thus the
associated exponent $\kappa$, has not yet been discussed. Our
analytical and numerical results for $\kappa$ thus provide new
insight into the geometry of DNA at melting, enabling one to make
more quantitative comparisons between the two types of approach.

\begin{acknowledgments}
Financial support by MIUR through COFIN 2001 and by INFM through
PAIS 2001 is gratefully acknowledged.
\end{acknowledgments}

\appendix
\section{The exponent $\mu$ for a self-avoiding loop}

The exponent $\mu$ may be evaluated for a self-avoiding loop by
generalizing the approach of McKenzie and Moore \cite{McKenzie}
who calculated this exponent for a self-avoiding walk. This
generalization closely follows the derivation in \cite{McKenzie}
and thus we will only briefly outline it here. The quantity of
interest is the probability distribution for two complementary
bases within a ring of length $l$. Equivalently this may be viewed
as the probability of two chains which are bound at one end, to
reach the same point $\vec{r}$. In this probability all possible
length $l_1$ and $l_2$ with $l_1+l_2=l$ are considered.

To this end we first consider the generating function of two
chains held together at one end and which are not restricted to
return to the same point $\vec{r}$
\begin{eqnarray}
&&\Gamma(\vec{r}_1,\vec{r}_2,\theta) =\nonumber \\
&&\sum_{l_1,l_2=1}^{\infty} \hspace{-0.3cm}C_{l_1,l_2}
P_{l_1,l_2}(\vec{r}_1,\vec{r}_2) s^{-(l_1+l_2)}
e^{-\theta(l_1+l_2)}
 \label{eq:green}
\end{eqnarray}
Here $C_{l_1,l_2}$ is the number of configurations of two chains
held together at one end, and $P_{l_1,l_2}(\vec{r}_1,\vec{r}_2)$
is the probability that the free end of one chain is at
$\vec{r}_1$ and the free end of the other chain is at $\vec{r}_2$.
The distribution function is given by
$\Gamma(\vec{r},\vec{r},\theta)$ where $\theta$ is a chemical
potential which controls the chain length $l_1+l_2$ in the sum
(\ref{eq:green}).

The Fourier transform of the Green's function (\ref{eq:green}) of
the two chains can be assumed to have the Ornstein and Zernike
form at small momenta $\vec{q}_1$ and $\vec{q}_1$,
\begin{equation}
\widehat{\Gamma}(\vec{q}_1,\vec{q}_2,\theta) \sim
\frac{\theta^{\nu
\rho}}{(\theta^{2\nu}+q_1^2)(\theta^{2\nu}+q_2^2)} \;.
\label{eq:qgreen}
\end{equation}
Moreover, since the two chains are bound together at one end their
total number of configurations is just that of one chain of length
$l_1+l_2$. Namely, $C_{l_1,l_2}=s^{l}l^{\gamma-1}$ where
$l=l_1+l_2$ and $l^{\gamma-1}$ is the usual enhancement factor for
a self-avoiding random walk. Using this in (\ref{eq:green}) one
can easily show that for small $\theta$
\begin{eqnarray}
\widehat{\Gamma}(0,0,\theta) &=& \sum_{l_1=1,l_2=1}^{\infty}
C_{l_1,l_2} s^{-(l_1+l_2)} e^{-\theta(l_1+l_2)} \nonumber \\
 &\sim& \theta^{-\gamma-1}
\end{eqnarray}
which after comparison with (\ref{eq:qgreen}), implies that
$\rho=4-(\gamma+1)/\nu$.

The quantity of interest is the probability
\begin{equation}
P_{l}(\vec{r},\theta)=\sum_{l_1+l_2=l}
P_{l_1,l_2}(\vec{r},\vec{r})
\end{equation}
This can be calculated by first inverting (\ref{eq:qgreen}) to
obtain $\Gamma(\vec{r_1},\vec{r_2},\theta)$. Setting
$\vec{r_1}=\vec{r_2}=\vec{r}$ yields
\begin{equation}
\Gamma(\vec{r},\vec{r},\theta) \sim \theta^{\nu(\rho+d-3)} r^{1-d}
e^{2 \theta^\nu r} \;. \label{eq:realgreen}
\end{equation}
One then has to carry out an inverse Laplace transform of
(\ref{eq:realgreen}) in order to extract $P_{l}(\vec{r},\theta)$.
One obtains
\begin{equation}
C_l P_{l}(\vec{r},\theta) s^{-l}= \frac{1}{2 \pi i}
\int_{X-i\pi}^{X+i\pi} d \theta e^{l \theta}
\Gamma(\vec{r},\vec{r},\theta) \;.
\end{equation}
where $C_l=C_{l_1,l_2}$ with $l=l_1+l_2$ and $X$ is larger than
the real part of any singularity of
$\Gamma(\vec{r},\vec{r},\theta)$. The result of this calculation
has the expected form (\ref{eq:dimu}) with
\begin{equation}
\mu=\frac{1}{1-\nu} \left( 1/2 + 2d(\nu-1/2)-\gamma \right) \;.
\end{equation}

\bibliography{biblio}

\begin{thebibliography}{28}
\expandafter\ifx\csname natexlab\endcsname\relax\def\natexlab#1{#1}\fi
\expandafter\ifx\csname bibnamefont\endcsname\relax
  \def\bibnamefont#1{#1}\fi
\expandafter\ifx\csname bibfnamefont\endcsname\relax
  \def\bibfnamefont#1{#1}\fi
\expandafter\ifx\csname citenamefont\endcsname\relax
  \def\citenamefont#1{#1}\fi
\expandafter\ifx\csname url\endcsname\relax
  \def\url#1{\texttt{#1}}\fi
\expandafter\ifx\csname urlprefix\endcsname\relax\def\urlprefix{URL }\fi
\providecommand{\bibinfo}[2]{#2}
\providecommand{\eprint}[2][]{\url{#2}}

\bibitem[{\citenamefont{Wartell and Benight}(1985)}]{wart85}
\bibinfo{author}{\bibfnamefont{R.~M.} \bibnamefont{Wartell}} \bibnamefont{and}
  \bibinfo{author}{\bibfnamefont{A.~S.} \bibnamefont{Benight}},
  \bibinfo{journal}{Phys. Rep.} \textbf{\bibinfo{volume}{85}},
  \bibinfo{pages}{67} (\bibinfo{year}{1985}).

\bibitem[{\citenamefont{Bonnet et~al.}(1998)\citenamefont{Bonnet, Krichevsky,
  and Libchaber}}]{bonn98}
\bibinfo{author}{\bibfnamefont{G.}~\bibnamefont{Bonnet}},
  \bibinfo{author}{\bibfnamefont{O.}~\bibnamefont{Krichevsky}},
  \bibnamefont{and}
  \bibinfo{author}{\bibfnamefont{A.}~\bibnamefont{Libchaber}},
  \bibinfo{journal}{Proc. Natl. Acad. Sci.} \textbf{\bibinfo{volume}{95}},
  \bibinfo{pages}{8602} (\bibinfo{year}{1998}).

\bibitem[{\citenamefont{Smith et~al.}(1992)\citenamefont{Smith, Finzi, and
  Bustamante}}]{DNA1}
\bibinfo{author}{\bibfnamefont{B.}~\bibnamefont{Smith}},
  \bibinfo{author}{\bibfnamefont{L.}~\bibnamefont{Finzi}}, \bibnamefont{and}
  \bibinfo{author}{\bibfnamefont{C.}~\bibnamefont{Bustamante}},
  \bibinfo{journal}{Science} \textbf{\bibinfo{volume}{258}},
  \bibinfo{pages}{1122} (\bibinfo{year}{1992}).

\bibitem[{\citenamefont{Strick et~al.}(1996)\citenamefont{Strick, Allemand,
  Bensimon, Bensimon, and Croquette}}]{DNA2}
\bibinfo{author}{\bibfnamefont{T.~R.} \bibnamefont{Strick}},
  \bibinfo{author}{\bibfnamefont{J.-F.} \bibnamefont{Allemand}},
  \bibinfo{author}{\bibfnamefont{D.}~\bibnamefont{Bensimon}},
  \bibinfo{author}{\bibfnamefont{A.}~\bibnamefont{Bensimon}}, \bibnamefont{and}
  \bibinfo{author}{\bibfnamefont{V.}~\bibnamefont{Croquette}},
  \bibinfo{journal}{Science} \textbf{\bibinfo{volume}{271}},
  \bibinfo{pages}{1835} (\bibinfo{year}{1996}).

\bibitem[{\citenamefont{L\'eger et~al.}(1999)\citenamefont{L\'eger, Romano,
  Sarkar, Robert, Bourdieu, Chatenay, and Marko}}]{DNA3}
\bibinfo{author}{\bibfnamefont{J.-F.} \bibnamefont{L\'eger}},
  \bibinfo{author}{\bibfnamefont{G.}~\bibnamefont{Romano}},
  \bibinfo{author}{\bibfnamefont{A.}~\bibnamefont{Sarkar}},
  \bibinfo{author}{\bibfnamefont{J.}~\bibnamefont{Robert}},
  \bibinfo{author}{\bibfnamefont{L.}~\bibnamefont{Bourdieu}},
  \bibinfo{author}{\bibfnamefont{D.}~\bibnamefont{Chatenay}}, \bibnamefont{and}
  \bibinfo{author}{\bibfnamefont{J.~F.} \bibnamefont{Marko}},
  \bibinfo{journal}{Phys. Rev. Lett.} \textbf{\bibinfo{volume}{83}},
  \bibinfo{pages}{1066} (\bibinfo{year}{1999}).

\bibitem[{\citenamefont{Cluzel et~al.}(1996)\citenamefont{Cluzel, Lebrun,
  Heller, Lavery, Viovy, Chatenay, and Caron}}]{DNA4}
\bibinfo{author}{\bibfnamefont{P.}~\bibnamefont{Cluzel}},
  \bibinfo{author}{\bibfnamefont{A.}~\bibnamefont{Lebrun}},
  \bibinfo{author}{\bibfnamefont{C.}~\bibnamefont{Heller}},
  \bibinfo{author}{\bibfnamefont{R.}~\bibnamefont{Lavery}},
  \bibinfo{author}{\bibfnamefont{J.-L.} \bibnamefont{Viovy}},
  \bibinfo{author}{\bibfnamefont{D.}~\bibnamefont{Chatenay}}, \bibnamefont{and}
  \bibinfo{author}{\bibfnamefont{F.}~\bibnamefont{Caron}},
  \bibinfo{journal}{Science} \textbf{\bibinfo{volume}{271}},
  \bibinfo{pages}{792} (\bibinfo{year}{1996}).

\bibitem[{\citenamefont{Smith et~al.}(1996)\citenamefont{Smith, Chui, and
  Bustamante}}]{DNA5}
\bibinfo{author}{\bibfnamefont{B.}~\bibnamefont{Smith}},
  \bibinfo{author}{\bibfnamefont{Y.}~\bibnamefont{Chui}}, \bibnamefont{and}
  \bibinfo{author}{\bibfnamefont{C.}~\bibnamefont{Bustamante}},
  \bibinfo{journal}{Science} \textbf{\bibinfo{volume}{271}},
  \bibinfo{pages}{795} (\bibinfo{year}{1996}).

\bibitem[{\citenamefont{Zimm}(1960)}]{zimm}
\bibinfo{author}{\bibfnamefont{B.~H.} \bibnamefont{Zimm}}, \bibinfo{journal}{J.
  Chem. Phys.} \textbf{\bibinfo{volume}{33}}, \bibinfo{pages}{1349}
  (\bibinfo{year}{1960}).

\bibitem[{\citenamefont{Poland and Scheraga}(1966{\natexlab{a}})}]{pola66}
\bibinfo{author}{\bibfnamefont{D.}~\bibnamefont{Poland}} \bibnamefont{and}
  \bibinfo{author}{\bibfnamefont{H.~A.} \bibnamefont{Scheraga}},
  \bibinfo{journal}{J. Chem. Phys.} \textbf{\bibinfo{volume}{45}},
  \bibinfo{pages}{1456} (\bibinfo{year}{1966}{\natexlab{a}}).

\bibitem[{\citenamefont{Fisher}(1966{\natexlab{a}})}]{fish66}
\bibinfo{author}{\bibfnamefont{M.~E.} \bibnamefont{Fisher}},
  \bibinfo{journal}{J. Chem. Phys.} \textbf{\bibinfo{volume}{45}},
  \bibinfo{pages}{1469} (\bibinfo{year}{1966}{\natexlab{a}}).

\bibitem[{\citenamefont{Poland and Scheraga}(1966{\natexlab{b}})}]{pola66b}
\bibinfo{author}{\bibfnamefont{D.}~\bibnamefont{Poland}} \bibnamefont{and}
  \bibinfo{author}{\bibfnamefont{H.~A.} \bibnamefont{Scheraga}},
  \bibinfo{journal}{J. Chem. Phys.} \textbf{\bibinfo{volume}{45}},
  \bibinfo{pages}{1464} (\bibinfo{year}{1966}{\natexlab{b}}).

\bibitem[{\citenamefont{Peyrard and Bishop}(1989)}]{peyr89}
\bibinfo{author}{\bibfnamefont{M.}~\bibnamefont{Peyrard}} \bibnamefont{and}
  \bibinfo{author}{\bibfnamefont{A.~R.} \bibnamefont{Bishop}},
  \bibinfo{journal}{Phys. Rev. Lett.} \textbf{\bibinfo{volume}{62}},
  \bibinfo{pages}{2755} (\bibinfo{year}{1989}).

\bibitem[{\citenamefont{Kafri et~al.}(2000)\citenamefont{Kafri, Mukamel, and
  Peliti}}]{kafr00}
\bibinfo{author}{\bibfnamefont{Y.}~\bibnamefont{Kafri}},
  \bibinfo{author}{\bibfnamefont{D.}~\bibnamefont{Mukamel}}, \bibnamefont{and}
  \bibinfo{author}{\bibfnamefont{L.}~\bibnamefont{Peliti}},
  \bibinfo{journal}{Phys. Rev. Lett.} \textbf{\bibinfo{volume}{85}},
  \bibinfo{pages}{4988} (\bibinfo{year}{2000}).

\bibitem[{\citenamefont{Kafri et~al.}(2002{\natexlab{a}})\citenamefont{Kafri,
  Mukamel, and Peliti}}]{kafr02}
\bibinfo{author}{\bibfnamefont{Y.}~\bibnamefont{Kafri}},
  \bibinfo{author}{\bibfnamefont{D.}~\bibnamefont{Mukamel}}, \bibnamefont{and}
  \bibinfo{author}{\bibfnamefont{L.}~\bibnamefont{Peliti}},
  \bibinfo{journal}{Eur. Phys. J. B} \textbf{\bibinfo{volume}{27}},
  \bibinfo{pages}{132} (\bibinfo{year}{2002}{\natexlab{a}}).

\bibitem[{\citenamefont{Kafri et~al.}(2002{\natexlab{b}})\citenamefont{Kafri,
  Mukamel, and Peliti}}]{kafr02B}
\bibinfo{author}{\bibfnamefont{Y.}~\bibnamefont{Kafri}},
  \bibinfo{author}{\bibfnamefont{D.}~\bibnamefont{Mukamel}}, \bibnamefont{and}
  \bibinfo{author}{\bibfnamefont{L.}~\bibnamefont{Peliti}},
  \bibinfo{journal}{Physica A} \textbf{\bibinfo{volume}{306}},
  \bibinfo{pages}{39} (\bibinfo{year}{2002}{\natexlab{b}}).

\bibitem[{\citenamefont{Duplantier}(1986)}]{dup1}
\bibinfo{author}{\bibfnamefont{B.}~\bibnamefont{Duplantier}},
  \bibinfo{journal}{Phys. Rev. Lett.} \textbf{\bibinfo{volume}{57}},
  \bibinfo{pages}{941} (\bibinfo{year}{1986}).

\bibitem[{\citenamefont{Duplantier}(1989)}]{dup2}
\bibinfo{author}{\bibfnamefont{B.}~\bibnamefont{Duplantier}},
  \bibinfo{journal}{J. Stat. Phys.} \textbf{\bibinfo{volume}{54}},
  \bibinfo{pages}{581} (\bibinfo{year}{1989}).

\bibitem[{\citenamefont{Causo et~al.}(2000)\citenamefont{Causo, Coluzzi, and
  Grassberger}}]{caus00}
\bibinfo{author}{\bibfnamefont{M.~S.} \bibnamefont{Causo}},
  \bibinfo{author}{\bibfnamefont{B.}~\bibnamefont{Coluzzi}}, \bibnamefont{and}
  \bibinfo{author}{\bibfnamefont{P.}~\bibnamefont{Grassberger}},
  \bibinfo{journal}{Phys. Rev. E} \textbf{\bibinfo{volume}{62}},
  \bibinfo{pages}{3958} (\bibinfo{year}{2000}).

\bibitem[{\citenamefont{Carlon et~al.}(2002)\citenamefont{Carlon, Orlandini,
  and Stella}}]{carl02}
\bibinfo{author}{\bibfnamefont{E.}~\bibnamefont{Carlon}},
  \bibinfo{author}{\bibfnamefont{E.}~\bibnamefont{Orlandini}},
  \bibnamefont{and} \bibinfo{author}{\bibfnamefont{A.~L.}
  \bibnamefont{Stella}}, \bibinfo{journal}{Phys. Rev. Lett.}
  \textbf{\bibinfo{volume}{88}}, \bibinfo{pages}{198101}
  (\bibinfo{year}{2002}).

\bibitem[{\citenamefont{Baiesi et~al.}(2002{\natexlab{a}})\citenamefont{Baiesi,
  Carlon, and Stella}}]{baie02}
\bibinfo{author}{\bibfnamefont{M.}~\bibnamefont{Baiesi}},
  \bibinfo{author}{\bibfnamefont{E.}~\bibnamefont{Carlon}}, \bibnamefont{and}
  \bibinfo{author}{\bibfnamefont{A.~L.} \bibnamefont{Stella}},
  \bibinfo{journal}{Phys. Rev. E} \textbf{\bibinfo{volume}{66}},
  \bibinfo{pages}{021804} (\bibinfo{year}{2002}{\natexlab{a}}).

\bibitem[{\citenamefont{Fisher}(1966{\natexlab{b}})}]{FisherSA}
\bibinfo{author}{\bibfnamefont{M.~E.} \bibnamefont{Fisher}},
  \bibinfo{journal}{J. Chem. Phys.} \textbf{\bibinfo{volume}{44}},
  \bibinfo{pages}{616} (\bibinfo{year}{1966}{\natexlab{b}}).

\bibitem[{\citenamefont{Redner}(2001)}]{redner}
\bibinfo{author}{\bibfnamefont{S.}~\bibnamefont{Redner}},
  \emph{\bibinfo{title}{A Guide to First-Passage Processes}}
  (\bibinfo{publisher}{Cambridge University Press}, \bibinfo{year}{2001}).

\bibitem[{\citenamefont{Lipowsky}(1991)}]{lipowsky}
\bibinfo{author}{\bibfnamefont{R.}~\bibnamefont{Lipowsky}},
  \bibinfo{journal}{Europhys. Lett.} \textbf{\bibinfo{volume}{15}},
  \bibinfo{pages}{703} (\bibinfo{year}{1991}).

\bibitem[{\citenamefont{Grassberger}(1997)}]{gras97}
\bibinfo{author}{\bibfnamefont{P.}~\bibnamefont{Grassberger}},
  \bibinfo{journal}{Phys. Rev. E} \textbf{\bibinfo{volume}{56}},
  \bibinfo{pages}{3682} (\bibinfo{year}{1997}).

\bibitem[{\citenamefont{Rosenbluth and Rosenbluth}(1955)}]{rose55}
\bibinfo{author}{\bibfnamefont{M.~N.} \bibnamefont{Rosenbluth}}
  \bibnamefont{and} \bibinfo{author}{\bibfnamefont{A.~W.}
  \bibnamefont{Rosenbluth}}, \bibinfo{journal}{J. Chem. Phys.}
  \textbf{\bibinfo{volume}{23}}, \bibinfo{pages}{356} (\bibinfo{year}{1955}).

\bibitem[{\citenamefont{Garel et~al.}(2001)\citenamefont{Garel, Monthus, and
  Orland}}]{gare01}
\bibinfo{author}{\bibfnamefont{T.}~\bibnamefont{Garel}},
  \bibinfo{author}{\bibfnamefont{C.}~\bibnamefont{Monthus}}, \bibnamefont{and}
  \bibinfo{author}{\bibfnamefont{H.}~\bibnamefont{Orland}},
  \bibinfo{journal}{Europhys. Lett.} \textbf{\bibinfo{volume}{55}},
  \bibinfo{pages}{132} (\bibinfo{year}{2001}).

\bibitem[{\citenamefont{Baiesi et~al.}(2002{\natexlab{b}})\citenamefont{Baiesi,
  Carlon, Orlandini, and Stella}}]{baie02b}
\bibinfo{author}{\bibfnamefont{M.}~\bibnamefont{Baiesi}},
  \bibinfo{author}{\bibfnamefont{E.}~\bibnamefont{Carlon}},
  \bibinfo{author}{\bibfnamefont{E.}~\bibnamefont{Orlandini}},
  \bibnamefont{and} \bibinfo{author}{\bibfnamefont{A.~L.}
  \bibnamefont{Stella}}, \bibinfo{journal}{Eur. Phys. J. B}
  \textbf{\bibinfo{volume}{29}}, \bibinfo{pages}{129}
  (\bibinfo{year}{2002}{\natexlab{b}}).

\bibitem[{\citenamefont{McKenzie and Moore}(1971)}]{McKenzie}
\bibinfo{author}{\bibfnamefont{D.~S.} \bibnamefont{McKenzie}} \bibnamefont{and}
  \bibinfo{author}{\bibfnamefont{M.~A.} \bibnamefont{Moore}},
  \bibinfo{journal}{J. Phys. A} \textbf{\bibinfo{volume}{4}},
  \bibinfo{pages}{L82} (\bibinfo{year}{1971}).

\end{thebibliography}

\end{document}